\documentclass[aps,prb,twocolumn,amssymb,amsmath,showpacs,superscriptaddress]{revtex4-1}
\usepackage{bm}
\usepackage{times}
\usepackage{graphicx}
\usepackage{color}
\usepackage{dcolumn}
\usepackage[colorlinks=true, letterpaper=true, pdfstartview=FitV, linkcolor=blue, citecolor=blue, urlcolor=blue]{hyperref}
\usepackage{appendix}
\usepackage[normalem]{ulem}

\newcommand{\vect}[1]{\boldsymbol{#1}}

\begin{document}
\title{Quantum intrinsic ${\cal T}$-odd spin Hall effect in altermagnets}

\author{Miaomiao Wei}
\thanks{These authors contributed equally to this work.}
\affiliation{College of Physics and Optoelectronic Engineering, Shenzhen University, Shenzhen 518060, China}
\author{Longjun Xiang}
\thanks{These authors contributed equally to this work.}
\affiliation{College of Physics and Optoelectronic Engineering, Shenzhen University, Shenzhen 518060, China}
\author{Fuming Xu}
\email[]{xufuming@szu.edu.cn}
\affiliation{College of Physics and Optoelectronic Engineering, Shenzhen University, Shenzhen 518060, China}
\affiliation{Quantum Science Center of Guangdong-Hongkong-Macao Greater Bay Area (Guangdong), Shenzhen 518045, China}
\author{Baigeng Wang}
\affiliation{National Laboratory of Solid State Microstructures and Department of Physics, Nanjing University, Nanjing 210093, China}
\affiliation{Collaborative Innovation Center for Advanced Microstructures, Nanjing 210093, China}
\author{Jian Wang}
\email[]{jianwang@hku.hk}
\affiliation{College of Physics and Optoelectronic Engineering, Shenzhen University, Shenzhen 518060, China}
\affiliation{Quantum Science Center of Guangdong-Hongkong-Macao Greater Bay Area (Guangdong), Shenzhen 518045, China}
\affiliation{Department of Physics, The University of Hong Kong, Pokfulam Road, Hong Kong, China}

\begin{abstract}

Drude weight, historically associated with the longitudinal Drude conductivity, can be generalized to describe the transverse or Hall component of the extrinsic conductivity tensor. In particular, transverse Drude weights, such as band geometric quantities Berry curvature dipole and spin vorticity, manifest themselves through the \textit{extrinsic} second-order nonlinear Hall effect and \textit{extrinsic} linear spin Hall effect (SHE) in diffusive transport, respectively. In this work, we uncover a new class of intrinsic Hall effects in quantum transport regime, termed as quantum intrinsic Hall effect (QIHE), which is the manifestation of system symmetry through intrinsic transport phenomena. For a given Hamiltonian, its transport characteristics can be revealed either intrinsically through QIHE in ballistic regime or extrinsically via the transverse Drude weight in diffusive transport, where both intrinsic and extrinsic effects share the same salient transport features governed by symmetry of the Hamiltonian. The physical origin of QIHE is attributed to quantum boundary scattering of the measurement setup that respects the system symmetry, as exemplified by the contact resistance of a two-terminal ballistic conductor. We demonstrate our finding by studying the quantum ${\cal T}$-odd ($\mathcal{T}$, time-reversal) SHE in altermagnets. Our work paves a way towards the quantum transport manifestation of band geometric characteristics.

\end{abstract}
\maketitle

\noindent \textit{Introduction} --- The Hall effects are closely related to local or global quantum geometry of Bloch electrons, such as Berry curvature.\cite{NiuRMP} Recently, various charge \cite{L-Fu,Guinea1,GaoQMD,Moore21,dxiao,syyang,M-Wei1,KTLawPRB,WeiPRL,SAYang23,XiangThird,Yan2024,Xiang24} and spin Hall effects \cite{spinHall2011,Sinova1,SHERev,Lee2018} driven by quantum geometry are proposed and some of them have been confirmed experimentally.\cite{MaBCDexp,KFMBCDexp,NHEBi2Se3,H-Yang,NatElec2021,NHEREV,XuSY2023,Gao2023,XiaoCMNHE,BPT2021,C-Song2,NingWang23} Among them, the extrinsic $\mathcal{T}$-odd ($\mathcal{T}$, time-reversal) spin Hall effect driven by the altermagnetic spin splitting effect (ASSE) has been observed in altermagnet RuO$_2$\cite{C-Song2}; the extrinsic $\mathcal{T}$-even charge nonlinear Hall effect (NHE) induced by Berry curvature dipole (BCD) has been detected in WTe$_2$\cite{MaBCDexp,KFMBCDexp} and other materials.\cite{NHEBi2Se3,H-Yang,NatElec2021,NHEREV} The experiments are performed on micron-sized diffusive samples and the corresponding Hall effects are classified as \textit{extrinsic} due to the dependence on the relaxation time $\tau$, which arises from impurity scattering within the semiclassical theory. The salient transport features of these Hall effects are governed by system symmetry and revealed by the Neumann's principle. For instance, in a $\mathcal{T}$-invariant system with single mirror symmetry, BCD-induced NHE has the optimal response when the driving electric field $\vect{E}$ is parallel to its BCD vector, whereas the Hall response vanishes if $\vect{E}$ is perpendicular to BCD. Nevertheless, in a ballistic system without impurity scattering, can these transport features be manifested via intrinsic effects?

On the other hand, to verify the Hall effects in experiment, one has to attach probing electrodes to the sample and build a measurement setup such as multi-probe Hall bars. The presence of electrodes has important impact on the transport response, since it introduces additional scattering and may hamper the system symmetry. It was discovered that disorder scattering in Pt electrodes can generate and transmit colossal NHE into the NbP sample that was not found to show NHE.\cite{MaoArxiv} The measurement setup is also crucial in quantum transport, which gives rise to boundary scattering in the ballistic regime and results in quantum effect such as contact resistance.\cite{Datta} When additional scattering between the electrodes and the sample (such as Schottky barrier) is properly removed\cite{MaBCDexp} and the measurement setup respects the sample symmetry, the transport response is solely determined by the Hamiltonian as well as quantum boundary scattering. As a result, if the extrinsic Hall effects discussed above can survive in ballistic conductors, how to describe their transport properties is beyond reach of the semiclassical theory and definitely needs to be addressed.

In this paper, we uncover a general class of intrinsic Hall effects in quantum transport regime, termed as quantum intrinsic Hall effect (QIHE), which is driven by band geometry as well as quantum boundary scattering that strictly respects the system symmetry. QIHE is the intrinsic manifestation of band geometry, and shares the same prominent transport features of the corresponding extrinsic effect which are described by transverse Drude weight in diffusive transport within the semiclassical theory. In the following, we first introduce the generalized Drude weight and the corresponding extrinsic Hall effect, then exemplify quantum boundary scattering in a two-terminal ballistic conductor, and finally demonstrate quantum intrinsic ${\cal T}$-odd SHE in altermagnets in the ballistic regime that displays the same transport properties of the extrinsic ${\cal T}$-odd SHE induced by ASSE. Intrinsic effects are more prominent, since they are not affected by the scattering process and directly related to band geometry.

\smallskip
\noindent{\it The extrinsic conductivity} --- We first discuss the physical origin of extrinsic effects. From the linear response theory, the frequency-dependent current is expressed as\cite{Moore1}
\begin{eqnarray}
\vect{J} = \sum_{mn} \int_k \dfrac{\vect{v}_{nm}
\left( f_{nm} \vect{\mathcal{A}}_{mn} + i\delta_{mn}\nabla_k f_m \right) \cdot \vect{E}(t) }{\omega-\epsilon_{mn}+i\eta}, \label{eq1}
\end{eqnarray}
where $f_{nm} = f_n - f_m$, $f_n$ is the Fermi distribution of band $n$, $\vect{\mathcal{A}}_{mn}$ is the interband Berry connection, $\vect{v}_{nm}$ is the velocity matrix element, $\epsilon_{mn} = \epsilon_{m} - \epsilon_{n}$ with $\epsilon_{n}$ the band energy, $\vect{E}(t)=\vect{E} \cos(\omega t)$, $\eta$ is an infinitesimal quantity to ensure the convergence at $t=-\infty$, and $\int_k = \int d\vect{k}/(2\pi)^d$ with $d$ the dimensionality of the system. The first term of Eq.~(\ref{eq1}) corresponds to the anomalous velocity $\vect{E} \times \vect{\Omega}$ with $\Omega$ the Berry curvature. The second term gives the Drude conductivity,
\begin{eqnarray}
\sigma^{D}_{xx} = \dfrac{i}{\omega+i\eta} \sum_n \int_k v_{n}^x
\partial_{x}f_n \equiv \dfrac{i}{\omega+i\eta} D^{xx}_1, \label{drude}
\end{eqnarray}
where $\partial_x \equiv \partial/\partial k_x$ and $D^{xx}_1=\sum_n \int_k v_{n}^x$ is the \textit{Drude weight}.\cite{Kohn} In the DC ($\omega \rightarrow 0$) and clean limit, the Drude conductivity is in general divergent, although the Drude weight is finite. However, since impurity scattering widely exists in diffusive conductors, one can introduce a phenomenological relaxation time $\tau$ to regulate this divergence, such that\cite{L-Fu}
\begin{align}
\tau \equiv \dfrac{1}{\eta}~ \Rightarrow ~ \sigma^{D}_{xx} = \tau D^{xx}_1.
\end{align}
Consequently, the Drude conductivity is an extrinsic quantity since it depends on $\tau$, which coincides with the semiclassical theory.\cite{L-Fu} Therefore, the Drude weight can only be observed through extrinsic effects. In the following, we generalize the Drude weight to describe extrinsic Hall effects.

\smallskip
\noindent \textit{Generalized Drude weight} --- The frequency-dependent conductivity is generally a complex quantity\cite{Kohn}
\begin{eqnarray}
\sigma(\omega) = \sigma'(\omega) + i\sigma''(\omega),
\end{eqnarray}
with $\sigma'$ ($\sigma''$) representing the real (imaginary) part. At the low frequency limit, the imaginary part $\sigma''$ diverges and the Drude weight was originally defined as\cite{Kohn,DrudeWeight}
\begin{eqnarray}
D = \pi \lim_{\omega \rightarrow 0} \omega \sigma''(\omega).
\end{eqnarray}
This definition can be generalized to include the transverse part of the conductivity. For example, the conductivity tensor of BCD-induced NHE is given by\cite{L-Fu}
\begin{align}
\sigma_{abc} = -\dfrac{i}{2 (\omega+i\eta)} \epsilon_{adc} \mathcal{D}_{bd} = -\dfrac{\tau}{2}\epsilon_{adc} \mathcal{D}_{bd}, \label{NHE}
\end{align}
where $\epsilon_{adc}$ is the Levi-Civita symbol and Einstein summation convention is adopted. Similar to Eq.~(\ref{drude}), BCD defined as
\begin{align}
\mathcal{D}_{bd}= \sum_n \int_k f_n (\partial_b \Omega_n^d) \label{BCD}
\end{align}
can be naturally viewed as the transverse Drude weight.\cite{Sodemann19} $\Omega^d_n$ is the Berry curvature of band $n$. Similarly, the $\mathcal{T}$-odd spin Hall conductivity is expressed as\cite{SVPRR}
\begin{align}
\sigma_{\alpha\beta}^{\gamma} = - \dfrac{i}{\omega+i\eta} \mathcal{S}_{\alpha\beta}^{\gamma} = -\tau \mathcal{S}_{\alpha\beta}^{\gamma}, \label{SHE0}
\end{align}
where the quantity $S_{\alpha\beta}^{\gamma}$
\begin{align}
\mathcal{S}_{\alpha\beta}^{\gamma} = \sum_n \int_k f_n \partial_a v_{n,\beta}^{\alpha}
\end{align}
can also be regarded as the transverse Drude weight, with $v_{n,\beta}^{\alpha}$ the spin velocity.\cite{SVPRR} Notably, the anti-symmetric part of $\mathcal{S}_{\alpha\beta}^{\gamma}$ is determined by the spin vorticity.\cite{note10} Both Eqs.~(\ref{NHE}) and (\ref{SHE0}) show that, the transverse Drude weights (BCD and spin vorticity) are manifested by extrinsic Hall effects in the DC limit.

Beyond the linear regime, we can further expand $\sigma''$ in terms of the Laurant series
\begin{eqnarray}
\sigma''(\omega) = \frac{D_1}{\pi \omega} + \frac{D_2}{\pi \omega^2} + ...
\end{eqnarray}
where the tensor $D_n$ with $n=1,2,...$ is the generalized Drude weight that consists of longitudinal (such as the second-order Drude conductivity\cite{Gao2021PRL}) and transverse components (such as Berry curvature multipole\cite{KTLawPRB}). Here $n$ labels the order of singular poles and the rank of $D_n$ depends on the order of electric fields in specific Hall effects. For instance, BCD is the off-diagonal part of the rank-3 Drude weight tensor while spin vorticity corresponds to the rank-2 spin Drude weight tensor. Similar to the conductivity tensor, the Drude weight tensor is fully determined by symmetry of the Hamiltonian, which in turn can probe band geometry through extrinsic effects.

\begin{figure}[tbp]
\includegraphics[width=\columnwidth]{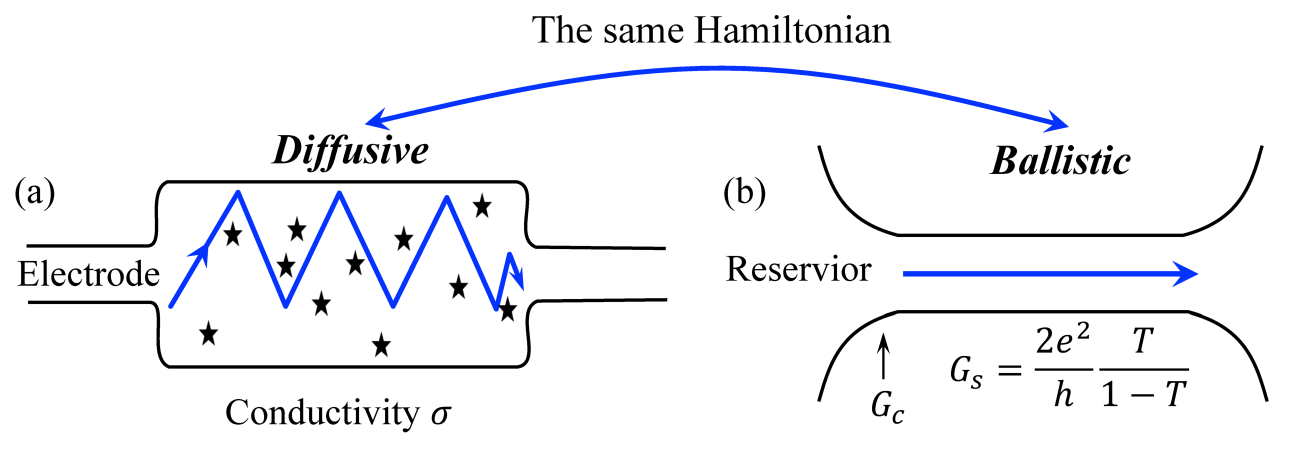}
\caption{(a) Schematic of a macroscopic conductor dominated by diffusive transport and characterized by the extrinsic conductivity $\sigma$. (b) Schematic of a mesoscopic conductor supporting ballistic transport and described by the intrinsic conductance $G = G_c + G_s$. Both (a) and (b) share the same Hamiltonian and hence the same symmetry, but they are in different transport regimes.}\label{Fig1}
\end{figure}

\smallskip
\noindent{\it Intrinsic versus extrinsic conductance} ---
In mesoscopic physics, a similar dilemma of diverging conductance puzzled the community for a long time. For the two-terminal mesoscopic conductor with single transmission channel as shown in Fig.~\ref{Fig1}(b), the conductance is intuitively given by\cite{Landauer1957,Landauer1970}
\begin{eqnarray}
G_{\rm s} = \frac{2e^2}{h} \frac{T}{1-T},
\end{eqnarray}
where $T$ is the transmission coefficient. Notice that $G_{\rm s}$ is the conductance of this conductor \textit{itself}. Clearly, in the case of perfect transmission $T=1$, the conductance will diverge and lead to zero resistance, which contradicts with experimental findings. There are two strategies to address this discrepancy: \\
(1) Imry et al. proposed that,\cite{Imry} in ballistic transport, the resistance of a two-terminal device consists of two parts:\cite{Imry,Datta}
\begin{eqnarray}
\frac{1}{G} = \frac{1}{G_{\rm c}} + \frac{1}{G_{\rm s}} = \frac{h}{2e^2} + \frac{h}{2e^2} \frac{1-T}{T} = \frac{h}{2e^2} \frac{1}{T}. \label{eqG}
\end{eqnarray}
Here $G_{\rm c}^{-1}$ is the contact resistance between the reservoir and the mesoscopic conductor, which is a quantum effect and arises from the measurement setup. In Eq.~(\ref{eqG}), although $G_{\rm s}$ is divergent for $T=1$, the total conductance is finite and recovers the conductance quanta $G = 2e^2/h$.\cite{Datta} This "regulation" from the measurement setup does not involve disorder scattering and hence leads to the \textit{intrinsic} conductance; \\
(2) One can also introduce disorder scattering into the conductor, then the perfect transmission is degraded to $T < 1$ and the divergence of $G_{\rm s}$ is removed. This mechanism is similar to the $\tau$-dependent extrinsic conductivity describing the situation in Fig.~\ref{Fig1}(a), which is referred to as the \textit{extrinsic} conductance. This extrinsic conductance shares the same transport characteristics of the intrinsic conductance. Notice that the macroscopic conductor in Fig.~\ref{Fig1}(a) and the mesoscopic conductor in Fig.~\ref{Fig1}(b) have the same Hamiltonian, but they are in different transport regimes and hence described by either extrinsic or intrinsic properties.

In summary, for quantum transport through a mesoscopic conductor, there are two kinds of scattering: quantum boundary scattering and disorder scattering. There are distinct differences between these scattering mechanisms. Quantum boundary scattering gives the finite intrinsic conductance while disorder scattering makes the conductance extrinsic. The contact resistance $G_{\rm c}^{-1}$ is one kind of quantum boundary scattering, which plays important roles in quantum transport. For instance, it was both experimentally\cite{Marcus92} and theoretically\cite{Bird97} found that, the magnetoconductance through a ballistic stadium-shaped quantum dot (QD) behaves distinctly from that of a circle-shaped QD. On the other hand, disorder scattering can drive the system into the diffusive regime where universal conductance fluctuation emerges, and even into the localized regime.\cite{RMT}

\begin{figure}[tbp]
\includegraphics[width=\columnwidth]{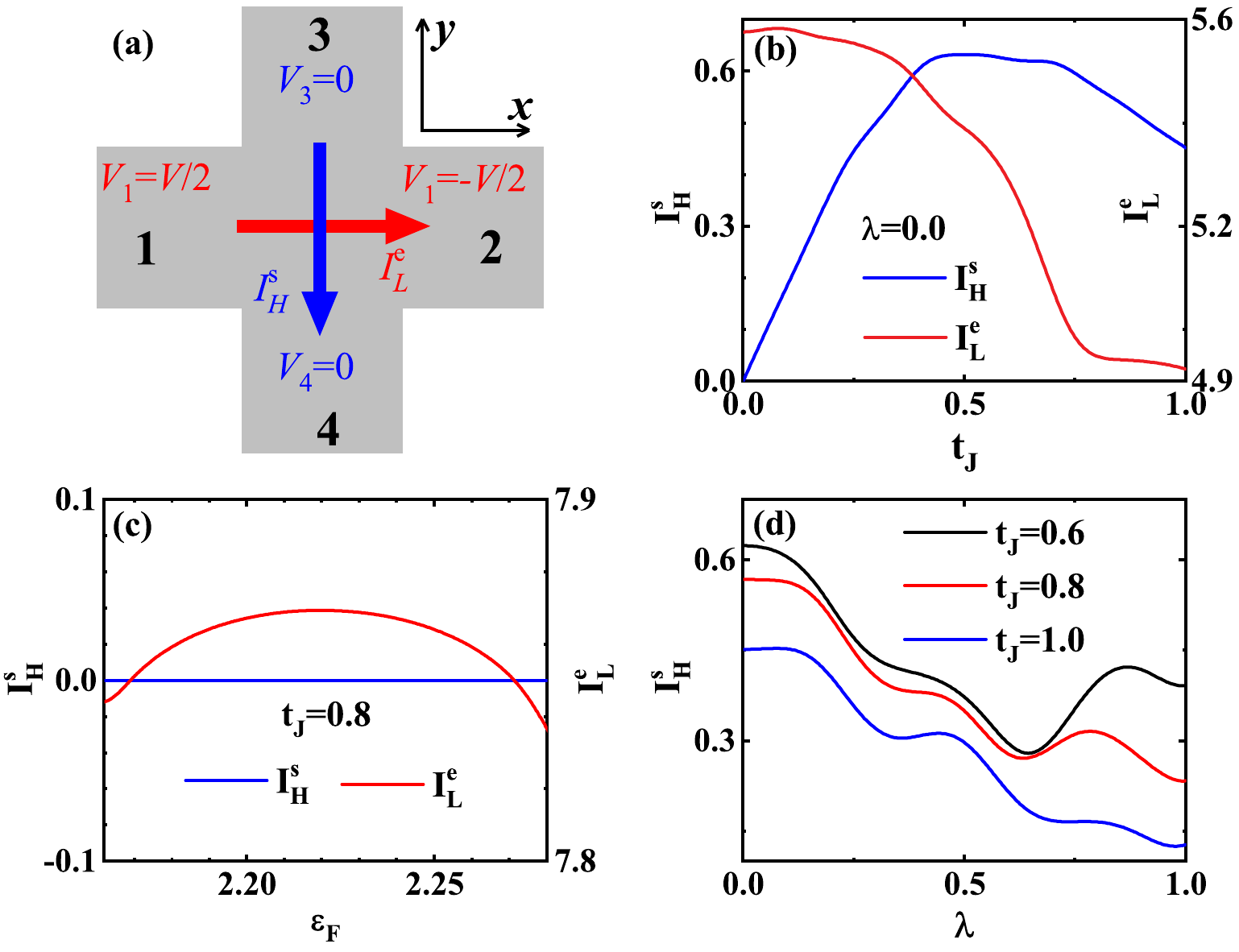}
\caption{(a) Schematic of the $\cal T$-odd SHE in a four-terminal system with closed boundary condition.\cite{boundcond,boundcond1,boundcond2} (b) Spin-Hall current $I^s_H$ and longitudinal charge current $I^e_L$ as a function of $t_J$. (c) Currents versus the Fermi energy $\varepsilon_F$ when the Hamiltonian is rotated by $45^o$. (d) $I^s_H$ as a function of the SOI strength $\lambda$ for different $t_J$. Parameter: $\varepsilon_F=2.2$ in (b) and (d).}\label{Fig2}
\end{figure}

Quantum boundary scattering also exists in multiterminal systems such as Hall bars and contributes to transport. If the sample is delicately prepared, where impurity scattering is absent and the measurement setup strictly respects the system symmetry, we deduce that intrinsic transport properties can be revealed in quantum transport regime. For instance, the BCD-induced NHE is known to be an extrinsic effect,\cite{L-Fu} as shown in Eq.~(\ref{NHE}). In Ref.~[\onlinecite{M-Wei}], a $\cal T$-invariant four-terminal system with single mirror symmetry was studied in the ballistic transport regime, similar to the setup in Fig.~{\ref{Fig2}}(a); as far as the symmetry is concerned, the observed second-order nonlinear Hall conductance has one-to-one correspondence with the BCD-induced extrinsic NHE that is obtained from the semiclassical theory for the same Hamiltonian. The system considered in Ref.~[\onlinecite{M-Wei}] is free of disorder, and the measurement setup or quantum boundary scattering respects the single mirror symmetry. Based on the above analysis, this example is a strong evidence of quantum intrinsic Hall effect (QIHE) driven by symmetry and quantum boundary scattering, which share the same salient transport feature as that of the BCD (transverse Drude weight)-induced extrinsic effect in diffusive transport. More importantly, this quantum intrinsic NHE was experimentally verified in high-quality graphene superlattices with low disorder scattering.\cite{NingWang23}

In the following, we study the spin Hall effect in altermagnets, and demonstrate that quantum intrinsic ${\cal T}$-odd SHE in a four-terminal device contains all characteristic features of its extrinsic counterpart derived from the semiclassical theory.

\smallskip
\noindent{\it Model Hamiltonian} ---
We start with the minimal two-band model that captures essential physics of altermagnets\cite{Sinova22}
\begin{eqnarray}
H = t k^2 + 2 t_J k_x k_y \sigma_z + \lambda (k_x \sigma_y - k_y \sigma_x), \label{ham}
\end{eqnarray}
where $t_J$ is the anisotropic exchange coupling constant and $\lambda$ is the Rashba spin-orbit interaction (SOI) strength. The second term breaks the time-reversal symmetry. Without loosing generality, we set $e=\hbar = t=1$ and assume $t_J < 1$ from now on. Turning off $t_J$, we have the conventional intrinsic ${\cal T}$-even SHE due to SOI\cite{Sinova}; when switching off $\lambda$, the altermagnetic phase emerges and the corresponding extrinsic ${\cal T}$-odd SHE driven by ASSE has been predicted theoretically and verified experimentally.\cite{Sinova1,Sinova2,C-Song1,C-Song2} When both $t_J$ and $\lambda$ are present, the Berry curvature is nonzero and the resulting SHE has both ${\cal T}$-odd and ${\cal T}$-even components.

\smallskip
\noindent{\it Extrinsic ${\cal T}$-odd SHE} --- We first set $\lambda=0$ and examine the extrinsic ${\cal T}$-odd SHE driven by ASSE using the semiclassical theory. The energy spectra of Eq.~(\ref{ham}) are given by $\varepsilon_\sigma = k^2 + 2 \sigma t_J k_x k_y$ with $\sigma = \uparrow, \downarrow$, where $\sigma_z$ is a good quantum number. According to the semiclassical theory, the current density is given by\cite{NiuRMP}
\begin{eqnarray}
{\bf J} = \int_k ({\bf v} + {\bf E} \times {\bf \Omega}) f.
\end{eqnarray}
Obviously, for Eq.~(\ref{ham}) the Berry connection is zero when SOI is absent and therefore transport due to ASSE is solely contributed from the Drude term. The linear longitudinal spin-resolved current density is found to be
\begin{eqnarray}
J^x_\sigma = -\tau \int_k (v^x_\sigma)^2 f'_0 E_x = \dfrac{ \tau \mu E_x}{\pi \sqrt{1-t_J^2}}, \label{temp1}
\end{eqnarray}
which is independent of the spin index. In Eq.~(\ref{temp1}), $f_0$ is the equilibrium Fermi distribution, $f'_0=\partial_\epsilon f_0$, and $\mu$ is the chemical potential.

For the transverse spin-resolved current density, we find\cite{note1}
\begin{eqnarray}
J^y_\sigma &=& -\tau \int_k v^x_\sigma v^y_\sigma f_0' E_x = \dfrac{\tau \mu \sigma t_J E_x}{2 \pi \sqrt{1-t_J^2}}, \label{temp3}
\end{eqnarray}
and similarly $J^x_\sigma = \tau \mu \sigma t_J E_y/( 2 \pi\sqrt{1-t_J^2} )$. Therefore, due to ASSE, the system generates a longitudinal charge current ${J}^e_L = {J}^x_\uparrow + {J}^x_\downarrow$ and a pure spin-Hall current ${J}^s_H = ({J}^y_\uparrow - {J}^y_\downarrow)/2$,\cite{Sinova1} as shown in Fig.~{\ref{Fig2}}(a). Both ${J}^e_L$ and ${J}^s_H$ are extrinsic effects due to the dependence on $\tau$. Different from the intrinsic $\mathcal{T}$-even SHE, the extrinsic SHE features a $\mathcal{T}$-odd behavior. This is because the relaxation time $\tau$ leads to $\mathcal{T} \sigma^{s}_{ab}=-\sigma^{s}_{ab}$ within the response relation $J_{a}^{s}=\tau \sigma_{ab}^s E_b$, in which both $J_a^s$ and $E_b$ are invariant under ${\cal T}$-symmetry whereas $\tau$ introduces a sign change.\cite{tauodd} For $\mathcal{T}$-odd SHE, the spin-Hall current $J_a^s$ should reverse its direction when the magnetization is switched. Indeed, for the altermagnetic Hamiltonian defined in Eq.~(\ref{ham}), when $t_J \rightarrow -t_J$, we find that $J_a^s \rightarrow -J_a^s$ from both the semiclassical result as well as quantum transport calculation shown below.

\smallskip
\noindent{\it Quantum intrinsic ${\cal T}$-odd SHE} ---
To explore quantum nature of the ${\cal T}$-odd SHE, we examine it in quantum transport regime using the Landauer-B\"{u}ttiker formula at zero temperature. For a multi-terminal system, the current in probe $\alpha$ is given by\cite{but-phys}
\begin{equation}
I_{\alpha} = \sum_\beta G_{\alpha\beta} V_\beta, \label{current1}
\end{equation}
where $G_{\alpha\beta} = {\rm Tr} [G^r \Gamma_\alpha G^a \Gamma_\beta]$ is the conductance tensor and $V_\alpha$ is the bias voltage of probe $\alpha$. The retarded Green's function is defined as $G^r(\varepsilon ) = 1/(\varepsilon-H-\Sigma^r)$ where $\Sigma^r = \sum_{\alpha} \Sigma^r_{\alpha}$ is the self-energy due to the probes and $\Gamma_\alpha=-2{\rm Im}\Sigma^r_\alpha$. The Hall and longitudinal currents are calculated under the closed boundary condition on a $N\times N$ square lattice with $N=20$ and $V=0.1$, as shown in Fig.~\ref{Fig2}(a).

Fig.~\ref{Fig2}(b) plots the pure spin-Hall current and longitudinal charge current versus the coupling constant $t_J$, where $I^s_H$ generally increases with $t_J$ whereas $I^e_L$ monotonically decreases. When rotating the Hamiltonian for $45^o$, we observe spin-polarized longitudinal charge currents with respect to $\varepsilon_F$ and zero spin-Hall current, as shown in Fig.~\ref{Fig2}(c). These phenomena capture all salient transport features of the extrinsic ${\cal T}$-odd SHE predicted theoretically in Ref.~[\onlinecite{Sinova1}]. However, in our quantum transport calculation, there is no disorder scattering and only quantum boundary scattering is present. Therefore, it corresponds to an intrinsic effect, which is termed as quantum intrinsic ${\cal T}$-odd SHE. In Fig.~\ref{Fig2}(d), we show the interplay between $t_J$ (ASSE) and $\lambda$ (SOI) on $I^s_H$. Clearly, both even and odd components of the quantum SHE exhibit similar transport behavior, and hence they all belong to intrinsic effects.

\begin{figure}[tbp]
\includegraphics[width=\columnwidth]{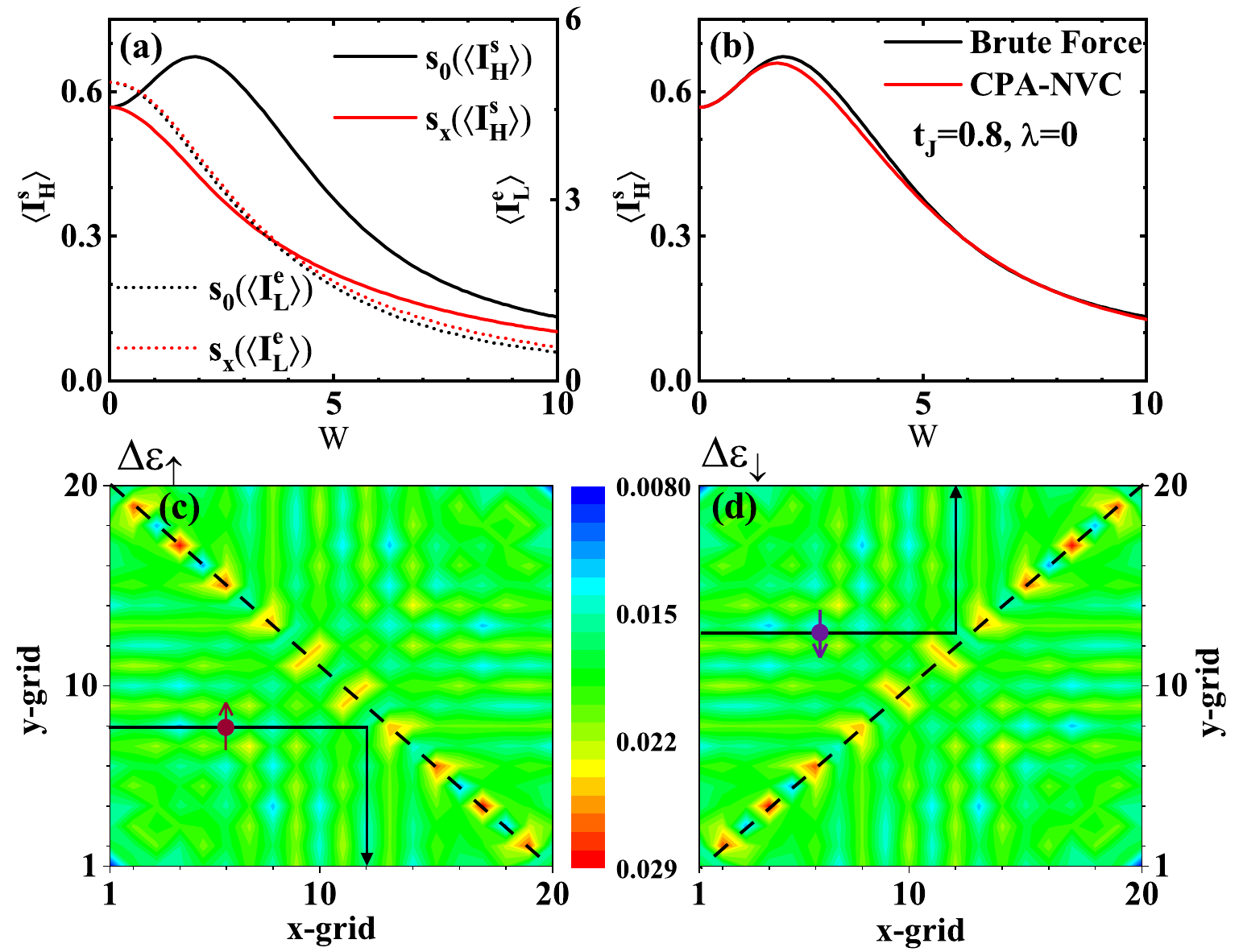}
\caption{(a) Average spin-Hall current and longitudinal charge current for different types of disorder ($s_0$: spin independent; $s_x$: spin flip). $\varepsilon_F=2.2$. (b) $\left< I^s_H \right>$ from different calculations. The black and red curves are, respectively, from the brute force and CPA-NVC methods. (c) and (d) Spin-resolved potential landscape $\Delta \varepsilon_\sigma$ due to disorder averaging at $W=2$.}\label{Fig3}
\end{figure}

\smallskip
\noindent{\it Disorder enhancement of ${\cal T}$-odd SHE}---
To study the disorder effect, we add on-site random potentials $V_r s_\alpha$ in the central region. $V_r$ is uniformly distributed in $[-W/2,W/2]$ with $W$ the disorder strength, and $s_\alpha$ is the Pauli matrix with $\alpha=0,x,y,z$. In Fig.~\ref{Fig3}(a), for spin-independent disorder, average spin-Hall current $\left< I^s_H \right>$ increases with $W$, reaches a peak at $W=2$ and then decreases exponentially to zero with the increasing of $W$. Clearly, $\left< I^s_H \right>$ demonstrates disorder-enhanced behavior. For spin-flip disorder $s_{x}$, the mixing among different spin channels tends to destroy the ASSE and therefore the enhancement of $\left< I^s_H \right>$ disappears. For the longitudinal charge current $\left< I^e_L \right>$, monotonic decreasing against $W$ is observed which is almost independent of disorder type. We find that, disorder enhancement of $\left< I^s_H \right>$ is a common feature occurring for a wide range of system parameters $t_J$ and $\varepsilon_F$.

To comprehend this enhancement behavior, we perform an analytic calculation using the coherent potential approximation (CPA) within nonequilibrium vortex correction (NVC).\cite{Y-Ke1,Y-Ke2,B-Fu} The CPA-NVC method produces an effective potential arising from disorder averaging, which can give insight to disorder-related phenomena. It was successfully applied to address the topological Anderson insulator.\cite{Beenakker09} Within CPA, the effective potential $\Delta \varepsilon$ is obtained through
\begin{eqnarray}
\langle G^r \rangle \equiv \frac{1}{\varepsilon + \Delta \varepsilon -H -\Sigma^r},
\end{eqnarray}
where $\langle G^r \rangle$ is the disorder-averaged Green's function. In Fig.~\ref{Fig3}(b), the results from CPA-NVC and brute force calculations are compared, and the agreement is surprisingly good for $\left< I^s_H \right>$. In Fig.~\ref{Fig3}(c) and (d), we plot the effective potential landscape $\Delta \varepsilon_\sigma$ generated by disorder averaging at $W=2$. For spin-up (-down) electrons coming from the left probe, there is a potential barrier along the dash line making $45^o$ ($135^o$) angle with the $x$-direction. Therefore, spin-up (-down) electrons tend to reach the down (top) probe but have to overcome the barrier to traverse to the top (down) or right probes, leading to the expected enhancement of spin-Hall current. Meanwhile, if the Hamiltonian is rotated by $45^o$, the potential barrier would be along $x$-direction. Clearly, in the presence of disorder, transport features of the setup remain the same. In another word, disorder-induced extrinsic effect shares the same transport characteristics of quantum intrinsic SHE.

\smallskip
\noindent{\it Conclusion} --- In conclusion, we have discovered a general class of quantum intrinsic Hall effects (QIHEs) in ballistic transport regime. QIHEs are driven by band geometry and quantum boundary scattering respecting the system symmetry, and exhibit the same transport features as their extrinsic counterparts induced by transverse Drude weights in diffusive regime. As a demonstration, quantum intrinsic ${\cal T}$-odd spin Hall effect in altermagnets is investigated in the absence and presence of disorder. Given the fact that ballistic transport evidence of ${\cal T}$-even SHE has been discovered in HgTe,\cite{Sinova10} quantum intrinsic ${\cal T}$-odd SHE can be revealed in a similar setup based on high-quality altermagnet nanostructures with ultra low disorder scattering. Therefore, experimental verification of our finding would be straightforward.

This work was supported by the National Natural Science Foundation of China (Grants No. 12034014 and No. 12174262).


\begin{thebibliography}{00}
\bibitem{note10}
The spin vorticity is defined as\cite{SVPRR}
\begin{align*}
w_n^\alpha
= \text{Re}\sum_{m \ne n} \dfrac{(\vect{v}_{nm} \times \vect{v}^\alpha_{mn})_z}{\epsilon_{nm}}
= \text{Im}\sum_{m \ne n} (\vect{v}^\alpha_{mn} \times \vect{\mathcal{A}}_{nm})_z.
\end{align*}
Note that by replacing $\vect{v}^\alpha_{mn}$ with Berry connection $\mathcal{A}_{mn}^\alpha$, this expression recovers the Berry curvature.
\bibitem{note1}
Note that in the limit of small $\Gamma \equiv 1/\tau$, the expression $\sigma^{\rm odd,a}_{bc} \sim {\rm Re}\int_k \sum_{mn}\frac{\Gamma^2 J^{a}_{b,mn} v^c_{nm}}{((\mu -\epsilon_n)^2+\Gamma^2)((\mu -\epsilon_m)^2+\Gamma^2)}$ given by Ref.~[\onlinecite{Sinova1}] recovers the result of Eq.~(\ref{temp3}).
\bibitem{boundcond}
The closed boundary condition refers to that: applying bias voltages on the probes as $V_1=-V_2=V/2$ and $V_3 = V_4 =0$, and measuring the Hall current $I_H = I_3 - I_4 $.\cite{boundcond1,boundcond2}
\end{thebibliography}
\end{document}